\documentclass[11pt]{article} 
\usepackage[utf8]{inputenc}
\usepackage{multirow}
\usepackage{amsfonts}
\usepackage{amsmath}
\usepackage{amssymb}
\usepackage{amsthm}
\usepackage{caption}
\usepackage[dvipsnames]{xcolor}
    \definecolor{darkgreen}{rgb}{0,0.5,0}
    \definecolor{darkblue}{rgb}{0,0,0.6}
    \definecolor{purple}{rgb}{0.4,.2,0.7}
\usepackage[margin = 2.5cm]{geometry}
\usepackage{graphicx}
\usepackage[hyperfootnotes = false, colorlinks = true, linkcolor = darkblue, citecolor = purple]{hyperref}
\usepackage{subcaption}
\usepackage{ytableau}

\usepackage{csquotes}

\usepackage{tikz}
\usetikzlibrary{decorations.pathmorphing}
\usetikzlibrary{decorations.markings}
\definecolor{mathred}{RGB}{180,44,37}
\definecolor{mathblue}{RGB}{39,94,190}
\tikzset{>=latex} 
\tikzset{ photon/.style={decorate, decoration={snake}, draw=black}}

\usepackage{comment}

\usepackage{physics}

\newcommand{\be}{\begin{equation}}
\newcommand{\ee}{\end{equation}}
\newcommand{\bea}{\begin{eqnarray}}
\newcommand{\eea}{\end{eqnarray}}

\def\tr{\mathrm{tr}}

\begin{document}

\thispagestyle{empty}
\begin{center}
    ~\vspace{5mm}

  \vskip 2cm 
  
   {\LARGE \bf 
       Bounding field excursions along null geodesics\\ with applications to cosmology
   }

   \vspace{0.5in}
     
   {\bf Aidan Herderschee${}^{a}$ and Aron C. Wall${}^{a, b}$
   }

    \vspace{0.5in}

  ${}^{a}$Institute for Advanced Study, Princeton, NJ 08540, USA\\
  ${}^{b}$DAMTP, Centre for Mathematical Sciences, \\
University of Cambridge, Wilberforce Road, Cambridge, U.K. 
                
    \vspace{0.5in}

    \vspace{0.5in}
    

\end{center}

\vspace{0.5in}

\begin{abstract}

Scalar fields in theories of gravity often inhabit a moduli space of vacua, and coherent spatial or temporal variations in their expectation values can produce measurable gravitational effects. Such variations are expected in contexts ranging from inflationary cosmology to the near-horizon regions of near-extremal black holes, where they can deflect light rays and shift horizons. This work derives a quantitative field excursion bound (FEB) on scalar variations along null geodesics, expressed in terms of the expansion parameter. The bound follows from the Raychaudhuri equation, assuming that all other fields satisfy the null energy condition (NEC).  It is saturated in certain spacetimes containing a timelike naked singularity. A possible generalization to semiclassical spacetimes that violate the NEC, but satisfy a strengthened version of the quantum focusing condition (QFC), is proposed. In cosmology, the FEB constrains the extent of large field excursions to be linearly bounded by the number of e-folds, independent of the inflationary model. This has notable implications for anthropic scenarios, where large excursions are often invoked to access favorable vacua.

\end{abstract}

\vspace{1in}

\pagebreak

\setcounter{tocdepth}{3}
{\hypersetup{linkcolor=black}\tableofcontents}

\section{Introduction}\label{sec:introduction}

Quantum field theories often possess a moduli space of vacua parameterized by the expectation values of scalar fields. When the expectation values vary coherently in spacetime, gravity couples and responds to these variations. Such variations are not merely an academic possibility: in the early universe, large-scale gradients in scalar field expectation values may have played a crucial role in inflation \cite{1983veu..conf..251S,Vilenkin:1983xq}, leaving imprints in the cosmic microwave background, while in extreme astrophysical environments, such as near-extremal black holes \cite{Ferrara:1995ih}, coherent moduli gradients can persist over macroscopic distances.\footnote{Arguments for Israel’s third law suggest that a (near-)extremal black hole cannot be produced by any finite, quasi-stationary physical process, thereby naively ruling out the astrophysical formation of such objects \cite{Israel:1986gqz}. However, arguments for Israel’s third law omit quantum effects; for example, Hawking evaporation can drive a magnetically charged black hole arbitrarily close to extremality on a timescale shorter than the age of the universe \cite{Maldacena:2020skw}. See also Ref. \cite{Kehle:2022uvc} for possible counterexamples to Israel's third law in classical contexts.} Even modest departures from uniformity can deflect light rays and shift horizons in spacetime. Taken together with the ongoing debate over whether distinct vacua in a gravitational theory belong to a single quantum theory \cite{Banks:2000zy,Banks:2003vp,Banks:2019oiz,Banks:2025nfe,Sen:2025bmj}, these effects underscore that such phenomena are a fundamental component of any realistic description of gravity–matter systems.

In this paper, we study bounds on field excursions in localized regions due to gravitational backreaction. An early indication of bounds on field excursions in localized regions appeared in Ref. \cite{Nicolis:2008wh}, which showed that trans-Planckian field configurations can be constructed within the Newtonian approximation, but only if the spatial extent of the experiment grows exponentially as a function of the excursion distance in the theory moduli space. Subsequent studies explored field variations in more phenomenological settings, including cosmic strings and bubbles of nothing \cite{Dolan:2017vmn, Draper:2019zbb, Draper:2019utz}. More recently, Ref. \cite{Delgado:2022dkz} demonstrated that regions deep in the vacuum manifold can be accessed using extremal black holes, whose entropy again grows exponentially with the field distance.\footnote{See also related work in Refs. \cite{Klaewer:2016kiy,Chaudhary:2020yyv,Sen:2025ljz,Sen:2025oeq}.}

Building on these results, we derive a quantitative bound, the field excursion bound (FEB), on field excursions along null geodesics, in terms of the expansion parameter $\theta$. The core intuition is that coherent variations in the expectation value of scalar fields induce gravitational focusing of light rays, and the expansion parameter $\theta$ quantifies the extent of this focusing. The Raychaudhuri equation \cite{Raychaudhuri:1953yv, Sachs:1961zz}, the scalar action, and the null energy condition (NEC) for all other fields enable us to derive the FEB. We verify the FEB in a spacetime containing a timelike naked singularity, where it is saturated. 

Our primary motivation for introducing the FEB comes from anthropic considerations in inflationary models of early-universe cosmology. In anthropic models of inflation, where the existence of observers guides vacuum selection, it is often assumed that scalar fields must traverse large distances in the theory moduli space to reach vacua with suitable low-energy properties. In cosmological settings, the FEB implies that the extent of large field excursions is linearly upper-bounded by the number of e-folds, independent of the underlying cosmological model. This leads to a striking implication: if an inflationary multiverse is indeed responsible for the observed fine-tuning of our universe, then the inflationary epoch should have lasted many e-folds. This prediction stands in significant tension with the Hartle-Hawking wavefunction \cite{Hartle:1983ai}, which favors initial conditions with minimal inflation \cite{Maldacena:2024uhs}. 

We also propose a potential generalization of the FEB to semiclassical spacetimes that violate the NEC. Our proposal relies on a strengthened Quantum Focusing Condition (QFC) \cite{Bousso:2015mna}. Although we do not prove the strengthened QFC, we note that it leads to some desirable properties of the spacetime beyond a quantum generalization of the FEB. Although the quantum FEB and strengthened QFC have no immediately evident phenomenological applications, they are nonetheless of significant formal interest and may prove relevant in future contexts.

The remainder of the paper is organized as follows. In Section \ref{febsec}, we derive a linear bound on field excursions under the assumption of the null energy condition. Section \ref{sec:nakedsingular} applies this bound to a spacetime containing a naked timelike singularity and demonstrates its saturation. In Section \ref{sec:largefieldinflat}, we turn to inflationary cosmology: we first analyze the translation-invariant case, then generalize the result to inhomogeneous inflationary models, and finally discuss the implications for anthropic phenomenology. Section \ref{quantumeff} extends our analysis to include quantum effects that violate the null energy condition. There, we conjecture a strengthened quantum focusing conjecture and use it to derive a quantum FEB. We conclude in Section \ref{disc} with a discussion of open questions and possible directions for future work.\\

\noindent \textbf{Notation}: We use mostly-plus signature for the metric, and we set Newton's constant $G_\text{N} = 1$, except in the quantum section where its renormalization is important. $G$ (with no $\text{N}$ subscript) refers to the determinant of the kinematic metric.

\section{The field excursion bound}\label{febsec}

In this section, we first give the field excursion bound (FEB), which we prove classically, using the standard form of the null energy for scalar fields (and assuming that all other matter fields obey the NEC). We are interested in field excursions along null geodesics of scalar theories with the action 
\begin{equation}\label{defdis}
S\ni \frac{1}{2}\int d^{D}x \sqrt{-g} \left [ G_{A,B}(\phi)\partial_{\mu}\phi^{A}\partial^{\mu}\phi^{B}+V(\phi)+\mathcal{L}_{\textrm{other}} \right ]\ .
\end{equation}
where $G_{A,B}(\phi)$ denotes the kinematic metric, $V(\phi)$ is a potential without derivative couplings, and $\mathcal{L}_{\textrm{other}}$ collects all additional contributions, including couplings between $\phi$ and other fields. We will assume that the contributions from $\mathcal{L}_{\textrm{other}}$ independently satisfy the NEC, which is further discussed in section \ref{disc}. 

To state the FEB, we first briefly review the geometry of null congruences.\footnote{See Appendix~F of Ref.~\cite{Carroll:2004st} for a detailed discussion.} 
Consider a family of null geodesics with an affinely parameterized, tangent vector field $k^{\mu}$. 
Introduce an auxiliary null vector $l^{\mu}$ such that both $k^{\mu}$ and $l^{\mu}$ are orthogonal to a codimension-two spatial hypersurface $\sigma$, whose evolution along the congruence we wish to track. 
We impose the conditions
\begin{equation}\label{eq:perpcond}
k^\mu \nabla_\mu k^\nu = 0, 
\quad l\cdot k = -1, 
\quad k^{\mu}\nabla_{\mu}l^{\nu} = 0 \ ,
\end{equation}
where the first ensures affine parametrization, the second fixes normalization, and the third specifies that $l^{\mu}$ is parallel transported along the geodesics. 
The induced metric on $\sigma$ is
\begin{equation}
h_{\sigma,\mu\nu} = g_{\mu\nu} + k_{\mu} l_{\nu} + l_{\mu} k_{\nu} \ ,
\end{equation}
and the expansion parameter,
\begin{equation}
\theta = h_{\sigma,\mu}^{\ \nu} \nabla_{\nu} k^{\mu} 
       = \frac{d}{d\lambda} \log\!\big(\sqrt{h_{\sigma}}\big) \ ,
\end{equation}
measures the fractional rate of change of the hypersurface area along the null congruence.

The FEB is 
\begin{displayquote}
FEB: Consider a
particular achronal null congruence with a null geodesic connecting two spacetime points $x_{1}$ and $x_{2}$. We denote the expectation values of the fields at these points $X_{1}$ and $X_{2}$.  Assuming the NEC, there is a bound on the difference between $X_{1}$ and $X_{2}$ of the form
\begin{equation}\label{eq:CFEB}
\left |\log\left ( \frac{\theta_{2}}{\theta_{1}}\right ) \right |\geq 4\sqrt{\frac{2\pi}{D-2}} d(X_{2},X_{1}) \ ,
\end{equation}
whenever $\text{sign}(\theta_1) = \text{sign}(\theta_2)$, where
$d(X_{2},X_{1})$ is the geodesic distance between $X_{2}$ and $X_{1}$ using the kinematic metric, $D$ is the spacetime dimension, and we are using units where $G = 1$.
\end{displayquote}
The intuition behind the FEB is that coherent variations in the matter fields cause light rays to focus and that $\theta$ quantifies the degree to which light rays focus. 
We assume the scalar fields are minimally coupled to gravity (otherwise, one must do a field redefinition to the Einstein frame before applying the FEB).

To prove the FEB, first consider a single scalar field with a standard kinetic term:
\begin{equation}\label{eq:scalarsource}
S=\frac{1}{16\pi}\int d^{D}x\sqrt{-g}R+\int d^{D}x \sqrt{-g} [\frac{1}{2}(\partial \phi(x))^{2}+V(\phi)+\mathcal{L}_{\textrm{other}}]
\end{equation}
where $R$ is the Ricci scalar, and $\mathcal{L}_{\textrm{other}}$ is the Lagrangian for the other matter fields.  This term may include arbitrary couplings to $\phi$ and $g_{ab}$, as long as the resulting stress-tensor contribution $T^{\mu\nu}_\textrm{other}$ from this term satisfies the NEC. 

The null Raychaudhuri equation \cite{Raychaudhuri:1953yv,Sachs:1961zz} implies that
\begin{equation}\label{eq:raychaudoriequ}
\frac{d\theta}{d\lambda}\leq \frac{-\theta^{2}}{D-2}-8\pi \left (  \frac{d\phi}{d\lambda}\right )^{2} \ ,
\end{equation}
where $D$ is the spacetime dimension, and we used Einstein's equation to identify
\begin{equation}\label{eq:einsteineqR}
R_{\lambda\lambda}\geq 8\pi \left (  \frac{d\phi}{d\lambda}\right )^{2} \ .
\end{equation}
The potential term $V(\phi)$ drops out of the Raychaudhuri equation because we are considering null geodesics. In Eq.~(\ref{eq:einsteineqR}), we are assuming that all other contributions to the null component of the stress-energy tensor from $\mathcal{L}_{\textrm{other}}$ are positive definite, i.e., they independently satisfy the NEC.

As a worst-case scenario for our bound, we consider minimizing 
\begin{equation}
\left | \frac{d\theta}{d\phi} \right | \ 
\end{equation}
along each point of the null geodesic.  (Because Eq. (\ref{eq:raychaudoriequ}) doesn't involve more than one derivative of $\theta$ or $\phi$, this minimization can be done independently at each point.)  We can divide each side of Eq. (\ref{eq:raychaudoriequ}) by $d\phi / d\lambda$ to obtain: 
\begin{equation}\label{eq:intermediatestep}
\left | \left (\frac{d\phi}{d\lambda}\right )^{-1}\frac{d\theta}{d\lambda} \right | \geq  \left | \frac{\theta^{2}}{D-2}\left (\frac{d\phi}{d\lambda}\right )^{-1}+8\pi \frac{d\phi}{d\lambda} \right | \ .
\end{equation}
The right-hand side is minimized when 
\begin{equation}
\left (\frac{d\phi}{d\lambda}\right )^{2}=\frac{\theta^{2}}{8\pi(D-2)}
\end{equation}
which implies the inequality  
\begin{equation}\label{eq:localexcurb}
\left | \frac{d\theta}{d\phi} \right | \geq 4 \sqrt{\frac{2\pi}{D-2}}  \theta \ .
\end{equation}
Integrating Eq. (\ref{eq:localexcurb}) along the null geodesic, one finds the FEB in Eq. (\ref{eq:CFEB}). The generalization to generic kinematic metrics is trivial. One could attempt to generalize the FEB to non-null geodesics, but then contributions from rigid sources would appear in the Raychaudhuri equation and the bound would be sensitive to the cosmological constant. 
\section{Example: naked time-like singularity}\label{sec:nakedsingular}

Field excursions can be arbitrarily large in the presence of time-like naked singularities and thereby provide a nice cross-check of the FEB. We consider the class of solutions given in Ref. \cite{Buchdahl:1959nk},\footnote{and further studied in Refs. \cite{Nicolis:2008wh}.} which considered an algorithm for constructing a solution with a non-trivial scalar field profile given a vacuum solution where the scalar field profile is trivial. The scalar is taken to be minimally coupled to gravity. Taking the Schwarschild solution as input, the algorithm outputs the metric
\begin{equation}\label{eq:singmetrci}
ds^{2}=-f^{\beta}dt^{2}+f^{-\beta}dr^{2}+r^{2}f^{1-\beta}d\Omega
\end{equation}
where
\begin{equation}\label{eq:mincoupledcalar}
\phi=\sqrt{\frac{(1-\beta^{2})}{16\pi}}\log(f), \quad f=1-\frac{2m}{r} \ .
\end{equation}
For $\beta\neq 1$, this metric exhibits a naked singularity at $r=2m$, where the event horizon would be. For simplicity, we again consider the Lagrangian in Eq. (\ref{eq:scalarsource}) with $J=0$.

We consider the tangent vector field, 
\begin{equation}\label{eq:nullgeosing}
k^{\mu}=(f^{-\beta},1,0,0)\ ,
\end{equation}
and the auxiliary vector, 
\begin{equation}
l^{\mu}=(\frac{1}{2},\frac{-f^{\beta}}{2},0,0) \ ,
\end{equation}
which both obey Eq. (\ref{eq:perpcond}). The spatial metric on the spacelike hypersurface is then 
\begin{equation}\label{eq:spatialme}
\begin{split}
h_{\sigma,\mu\nu}&=g_{\mu\nu}+l_{\mu}k_{\nu}+l_{\nu}k_{\mu}  \\
&=\begin{bmatrix}
0 & 0 & 0 &  0\\
0 & 0 & 0 &  0\\
0 & 0 & f^{1-\beta}r^{2} & 0 \\
0 & 0 & 0 & f^{1-\beta}r^{2}\sin^{2}(\theta)
\end{bmatrix} \ .
\end{split}
\end{equation}
Using Eqs. (\ref{eq:nullgeosing}) and (\ref{eq:spatialme}), we compute the expansion parameter
\begin{equation}
\begin{split}
\theta=\frac{2 (-m\beta +r-m)}{(r-2m) r}\ .
\end{split}
\end{equation}
Note that the expansion parameter diverges at $r=2m$ unless $\beta=1$ \cite{Carroll:2004st}. To cross-check the FEB, we want to check the FEB at every point along a geodesic. We therefore consider the infinitesimal change in $\log(\theta)$ relative to $\phi$:
\begin{equation}
\begin{split}
\left | \frac{d\log(\theta)}{d\phi}\right|&=\left | \frac{d\log(\theta)}{dr} \left ( \frac{d\phi}{dr} \right )^{-1}\right |\\
&=\left | \frac{2 \sqrt{\pi } \left(2 (\beta +1) m^2-2 (\beta +1) m r+r^2\right)}{\sqrt{1-\beta ^2} m (\beta  m+m-r)} \right | \ .
\end{split}
\end{equation}
The above function is minimized at 
\begin{equation}
\left | \frac{d\log(\theta)}{d\phi}\right||_{r=r_{\textrm{min}}}=4\sqrt{\pi}, \quad r_{\textrm{min}}=m(1+\beta \pm \sqrt{1-\beta^{2}}) \ ,
\end{equation}
which corresponds to saturation of the FEB with $G=1$ at $r_{\textrm{min}}$.

\section{Example: large-field inflation}\label{sec:largefieldinflat}

In this section, we consider the FEB in the context of inflation. Inflation is currently the leading model of early universe cosmology as it explains the horizon and flatness problems \cite{Guth:1980zm,Linde:1981mu,Albrecht:1982wi}.\footnote{See Ref. \cite{Baumann:2009ds} for a nice review.} The general idea is that there is a field called the inflaton that transversed a tall, flat potential, $V(\phi)$. Due to the high potential energy of the inflation during this period, the universe was rapidly expanding and de-Sitter-like. Eventually, the inflaton fell into a well in $V(\phi)$, which led to reheating and the near-flat cosmology we are familiar with. 

\subsection{Translation-invariant case}
\label{sec:transinvar}
We first assume the metric before and during inflation took the form 
\begin{equation}\label{eq:inflation}
ds^{2}=-dt^{2}+a(t)^{2}[dx^{2}+dy^{2}+dz^{2}] \ ,
\end{equation}
where the Hubble parameter, $H(t)$, and number of e-folds, $\Delta N$, were given by 
\begin{equation}\label{def}
H(t)=\frac{\partial_{t}a(t)}{a(t)}, \quad \Delta N=\log (\frac{a(t)}{a(t_{0})}) \ .
\end{equation}
where $t_{0}<t$. We consider a generic $H(t)$ because there is no experimental evidence that $H(t)$ was even approximately constant more than 30 e-folds before reheating. We again consider a minimally coupled scalar field as in Eq. (\ref{eq:scalarsource}). 

We first assume translational invariance. Consider a null congruence defined by the tangent vector field
\begin{equation}
k^{\mu}=(\frac{1}{a(t)},\frac{1}{a(t)^{2}},0,0)\ .
\end{equation}
and auxiliary vector $l^{\mu}$,
\begin{equation}
l^{\mu}=(\frac{a(t)}{2},\frac{1}{2},0,0) \ ,
\end{equation}
which again obey Eq. (\ref{eq:perpcond}). The resulting metric for the spatial hypersurface is 
\begin{equation}
\begin{split}
h_{\sigma,\mu\nu}=\begin{bmatrix}
0 & 0 & 0 &  0\\
0 & 0 & 0 &  0\\
0 & 0 & a(t)^{2} & 0 \\
0 & 0 & 0 &  a(t)^{2}
\end{bmatrix}
\end{split}
\end{equation}
and the expansion parameter is  
\begin{equation}\label{eq:theta}
\begin{split}
\theta&=\frac{\partial_{t}a(t)}{a(t)^{2}}   \ .
\end{split}
\end{equation}
Plugging Eq. (\ref{eq:theta}) into the FEB yields
\begin{equation}\label{eq:efoldboundlin}
4\sqrt{\pi}|\phi_{0}-\phi| \leq \Delta N + \log(\frac{H(t_0)}{H(t)}).
\end{equation}
The logarithmic term should be subleading if we assume that inflation is in the slow-roll approximation for most of the change in $\Delta \phi$. In that case, the bound on the number of e-folds is effectively
\begin{equation}\label{eq:efoldbound}
\Delta N \gtrsim 4\sqrt{\pi}|\phi_{0}-\phi| \ .
\end{equation}
Therefore, we find that field excursions along null geodesics are linearly bounded by the number of e-folds.

In deriving Eq.~(\ref{eq:efoldbound}), we omitted the term $\log\big(H(t_{0})/H(t)\big)$. However, this contribution is nonnegative. The NEC implies that $H(t)$ is decreasing with time, so the logarithmic term can only weaken the bound. While tilt parameters constrain the fractional time variation of $H$ within the observable window of inflation (roughly $30$–$60$ e-folds) \cite{Liddle:2000cg}, they provide little information about $H(t)$ at earlier times, leaving open the possibility that the logarithmic term could become relevant. For example, suppose that
\begin{equation}
\Delta N = C \log\!\left(\frac{H(t_{0})}{H(t)}\right)
\end{equation}
for some constant $C$. From Eq.~(\ref{def}), the corresponding solution for the scale factor is
\begin{equation}
a(t) = (C t)^{1/C} \, ,
\end{equation}
which describes FRW metrics with flat spatial slices, where the power-law exponent 
depends on the dominant matter content. In such cases, the bound in Eq.~(\ref{eq:efoldbound}) can be weakened. However, because the logarithmic and $\Delta N$ terms are proportional, only the numerical coefficient in Eq.~(\ref{eq:efoldbound}) is altered, remaining $\mathcal{O}(1)$, so there is still a linear bound on $|\phi_{0}-\phi|$ in terms of $\Delta N$. In general, the logarithmic term becomes more relevant in Eq.~(\ref{eq:efoldbound}) the slower that $a(t)$ grows with $t$. Therefore, to obtain spacetimes in which the logarithmic term dominates parametrically over 
the $\Delta N$ contribution, so that no linear field-excursion bound exists, the 
scale factor would need to grow slower than any power-law, for instance,  
\begin{equation}
a(t) = \alpha \log\!\left(\tfrac{t}{t_{0}}\right) \, .
\end{equation}
However, such behavior would require exotic matter content and is unlikely to be relevant 
for realistic cosmological models. Even allowing power-law behavior of $a(t)$ is a very conservative assumption, since a large cosmological constant is expected during the inflationary epoch, whereas power-law behavior for $a(t)$ requires the cosmological constant to vanish.

\subsection{Less symmetrical inflation}

In the preceding section, we assumed a metric with spatial translational symmetry, but this restriction is unnecessary. Intuitively, this is because the FEB can be defined for a single null geodesic.  We now provide an analysis valid for generic spacetimes. The goal is to derive a bound on the local field excursions of a scalar in terms of a generalized notion of e-folds, applicable along a single light ray.

\paragraph{Preliminary Assumptions.}  Let us consider two specific Cauchy slices, an initial slice $\Sigma_i$ near the start of inflation, and a final slice $\Sigma_f$ near the end of inflation.  For simplicity, we assume that $D = 4$ throughout, even in the deep inflationary past.  The goal is to place a bound on the maximum excursion of the scalar field $\phi$ from $\Sigma_i$ to $\Sigma_f$, expressed in terms morally similar to eq \eqref{eq:efoldbound}, appropriately generalizing the redshift factor $\Delta N$ and the Hubble ratio $H_f / H_i$.

\begin{figure}[t]
    \centering
\begin{tikzpicture}
  \node[anchor=center,inner sep=0] at (0,0) 
    {\includegraphics[width=12cm]{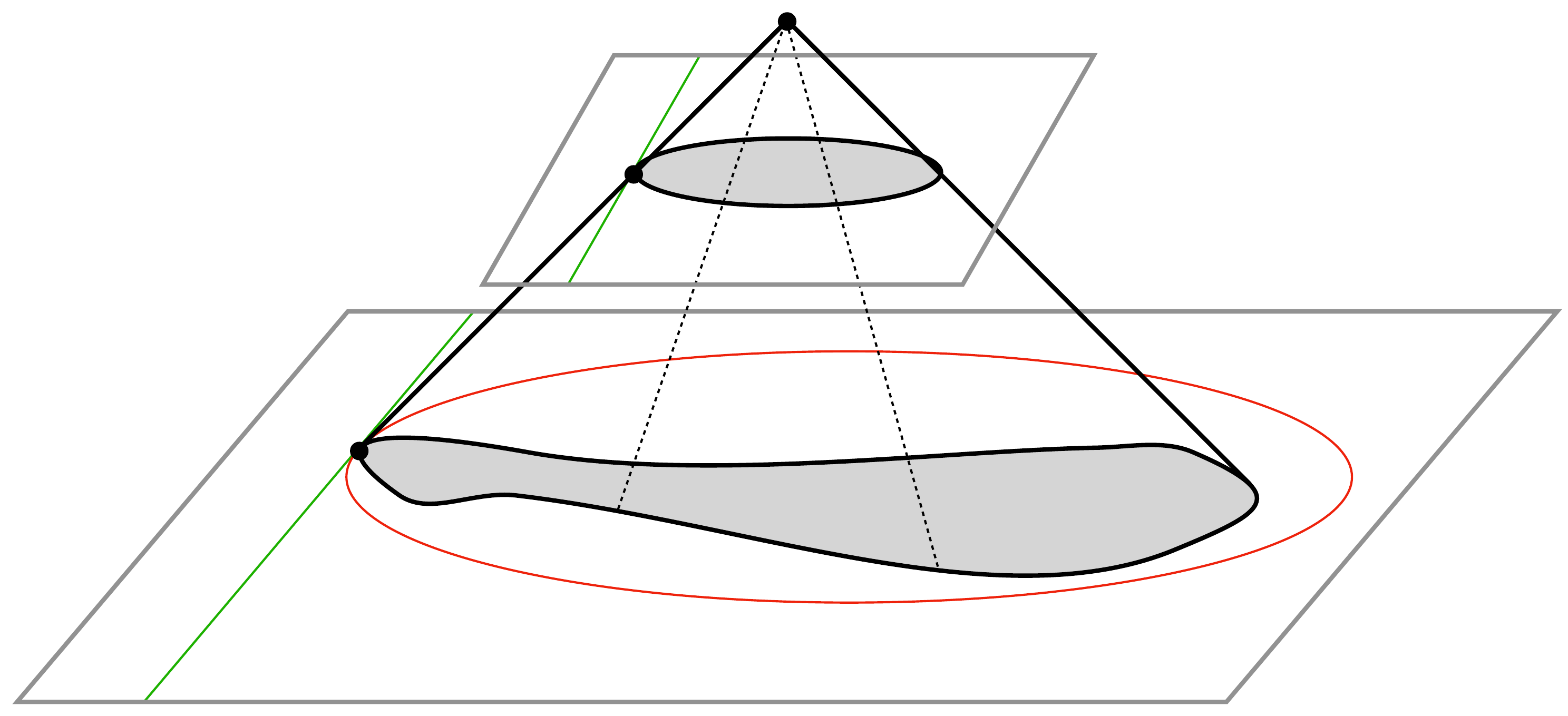}};
\node[text=green] at (-4.7,-2) {$s_{i}$};
\node at (-2.3+0.05,0.1-0.05) {$\ell$};
\node[text=green] at (-1.1,1.9) {$s_{f}$};
\node at (-3.45,-0.6) {$q$};
\node at (-2.3,0) {};
\node[text=red] at (0,-2.1) {$\kappa_{i}$};
\node at (-1.1,-1.45) {$\gamma_{i}$};
\node at (0.8,1) {$\gamma_{f}$};
\node at (1.7,1.9) {$\Sigma_{f}$};
\node at (5,0) {$\Sigma_{i}$};
\node at (0,5/2+0.3) {$p$};
\end{tikzpicture}
\caption{Schematic depiction of the initial and final time slices, denoted by $\Sigma_{i}$ and $\Sigma_{f}$, respectively. The cone corresponds to the past lightcone of $p$, which is denoted as $P$.} \label{pictoralargument}
\end{figure}

Let us trace back in time to the past $P = J^{-}(p)$ of a point $p$, representing present-day observations from Earth, as shown in Fig. \ref{pictoralargument}.  Our past light-cone $\partial P$ consists of light rays shot out from Earth at various solid angles.  If gravitational lensing causes these light rays to intersect, the light rays generating 
$\partial P$ will be truncated at these intersection points; hence $\partial P$ is an achronal surface (i.e.~no two points are timelike separated).  We define $\gamma_i = \partial P \cap \Sigma_i$, and $\gamma_f = \partial P \cap \Sigma_f$.  These are 2-dimensional surfaces.\footnote{Assuming that spacetime is globally hyperbolic and noncompact, they should be nonempty surfaces, that divide each $\Sigma$ into an interior (points in our past $P$) and exterior (points not in our past $P$).}

Given the homogeneity and isotropy of the standard Big Bang model, we may (to a high degree of accuracy) regard the geometry of $\Sigma_f$ as being that of flat Euclidean 3-space, with a determinate Hubble constant $H_f$.  $\gamma_f$ is thus a 2-sphere, of very large radius compared to $1/H_f$.

We do not, however, wish to assume that the spacetime manifold $\cal M$ is homogeneous or isotropic at still earlier times, at $\Sigma_i$ and the time periods between $\Sigma_i$ and $\Sigma_f$.  This means in particular that $\phi$ need not be constant on $\Sigma_i$.  Furthermore, the geometry of $\Sigma_i$ need not be constant; instead, it is defined by a pair $(q_{ab}, K_{ab})$.  Here $q_{ab}$ is the pullback of the metric to $\Sigma_i$, with $a,b \ldots \in \{1,2,3\}$ indices restricted to lie in the $\Sigma_i$ slice.  And $K_{ab} = \frac{1}{2} dq_{ab}/dt$ is the extrinsic curvature of $\Sigma_i$ within $\cal M$, where $t$ represents the unit normal time direction.  If the universe is everywhere expanding on the slice $\Sigma_i$ (albeit possibly in a non-isotropic manner), then we will have $K_{ab} \ge 0$ as a matrix.

We will need to make one substantive assumption on the geometry, which is this:

\vspace{10pt}
\noindent \textbf{Mean Convexity Assumption (MCA):} $\gamma_i$ is \emph{marginally bounded} within a \emph{mean-convex} closed region $R \in \Sigma_i$.
\vspace{10pt}

\noindent By ``marginally bounded'', we mean that $\gamma_i$ is contained within the set $R$, but also touches $R$ on at least one point of its boundary $\kappa_i = \partial R$.  By ``mean-convex'', we mean that $\kappa_i = \partial R$ satisfies a positivity condition on the trace of its outward-pointing extrinsic curvature: $k_{ij} h^{ij} \ge 0$.\footnote{This is a weaker assumption than local convexity, which would assert that $k_{ij} \ge 0$ as a matrix.}  Here $h_{ij}$ is the pullback of the metric to $\kappa_i$, with tangent indices $i,j \in \{1,2\}$, and $k_{ij} = \frac{1}{2} dq_{ij}/dn$ is the extrinsic curvature of $\kappa_i$ within $\Sigma_i$, and $n$ is the outer normal direction.  (We use lowercase $k_{ij}$ to distinguish this from the timelike extrinsic curvature $K_{ab}$ defined previously.) 

The purpose of the MCA is to rule out counterexamples like the following: let $\Sigma_i = S^3$ have the geometry of a 3-sphere, and let our past $P \cap \Sigma_i$ be everything except a very small ball $B \in \Sigma_i$ of some radius $r$.  In this case, even if we assume uniformity and isotropy (resulting in some uniform Hubble constant $H_i > 0$), the expansion $\theta_i[\gamma_i = \partial B]$ can be arbitrarily large, and diverges as $r \to 0$.  In this particular example, the MCA requires that our past $P$ be confined to at most one hemisphere of the 3-sphere $\Sigma_i$.\footnote{If the geometry of our universe is close to global de Sitter, then our past $P$ will occupy much more than one hemisphere in the contracting phase of de Sitter, when $H < 0$.  However, we do not intend for our results to apply to slices $\Sigma_i$ in such a contracting phase, so this is not relevant to the discussion here.}  This seems physically realistic given a standard inflationary scenario, since in such cases we expect that some fairly small region $R$ will grow exponentially in a way that causally decouples it from the rest of the universe.  Thus, it can easily contain within it our entire past.\footnote{If this assumption is not true at some time $t_1$, we can always wait a few e-folds to a new time $t_2$, and then take $\Sigma_i = t_2$.  The only cost of doing so is that our excursion bound on $\phi$ will only go back to time $t_2$ instead of to time $t_1$, but our excursion bound will still cover the bulk of the time spent during inflation.}

On the other hand, the MCA is automatically satisfied if the geometry of $\Sigma_i$ is noncompact and ``sufficiently close'' to the geometry of flat Euclidean 3-space (assuming that the spacetime $\cal M$ is globally hyperbolic, so that $\gamma_i$ remains compact).  $\gamma_i$ is thus marginally bounded by some 2-sphere $\kappa_i$ of constant radius $r = r_\kappa$, where $r$ is the standard radial coordinate of Euclidean 3-space, around some arbitrarily selected origin.  (This marginally bounding sphere may be obtained by starting with a large sphere and shrinking it, stopping when it first touches $\gamma_i$.)  Such spheres have a mean-curvature of $k_{ij}h^{ij} = 2/r_\kappa > 0$.  As this is strictly positive, any sufficiently mild deformation of the geometry should preserve this condition.  The same result holds for small perturbations to a hyperbolic 3-geometry.

Although these examples are sufficient for the MCA to be satisfied, it does not in any way require the geometry of $\Sigma_i$ to be close to uniform.  We only need to assume that the geometry is not \emph{so} strongly distorted, that moving outward \emph{decreases} the area element.  This might well not be true at the very beginning of the universe, when it is tiny and possibly highly curved.  But in general, inflation has the tendency to flatten out the spatial metric, so we do expect that this property will be true a short time after the beginning of inflation, if not before.

\paragraph{Construction of Lightray.}

Given our assumption of the MCA, it follows that $\gamma_i$ touches the mean-convex surface $\kappa_i$ on at least one point $q$.  Let $\ell$ be the lightray which extends from our present $p$ back to this point $q \in (\gamma_i \cap \kappa_i) \in \Sigma_i$.  By construction, $\ell$ is achronal.

\paragraph{Definition of Angular Redshift.}

The lightray $\ell$ has an affine parameter $\lambda$, which may be used to define an (angular) redshift factor between the slices $\Sigma_i$ and $\Sigma_f$.  (This is the same as the redshift that would actually be experienced physically, by a test photon propagating along $\ell$.)  Specifically, for any Cauchy slice $\Sigma$ with unit normal $t$, we define the redshift factor as follows:
\begin{equation}\label{redshift}
1+z = \frac{dt}{d\lambda},
\end{equation}
where the affine parameter $\lambda$ is normalized so that $z = 0$ at the present-day $p$.\footnote{Here we are acceding to the conventional definition of the redshift using $1+z$ instead of $z$, so that $z = 0$ at the present day.  However, it should be noted that during inflationary times, $z \gg 1$, so the +1 is a small error that can be neglected.}  However, the normalization will drop out of our final results, because we are only interested in the effective number of e-folds during inflation, which we define as the log ratio of the redshift factor between $\Sigma_i$ and $\Sigma_f$:
\begin{equation}\label{eq:deltaNinhomo}
\Delta N = \log\left (\frac{1+z_{f}}{1+z_{i}}\right ) 
= \log\left(\frac{(dt/d\lambda)_f}{(dt/d\lambda)_i}\right) ,
\end{equation}

\paragraph{Definition of Angular Hubble Constant.}

Next, we provide a local definition of the Hubble constant $H$ on a Cauchy slice $\Sigma$.  Let $n^a$ be any unit tangent vector on any point $q$ of $\Sigma$.  Then, half the local expansion of the area element perpendicular to $n^a$ is given by:
\begin{equation}
H(q,n^a) = \frac{1}{2} K_{ab}(q^{ab} - n^a n^b).
\end{equation}
This provides an (angle- and position-dependent) definition of the Hubble constant $H$, which is independent of the sign of the unit vector $n^a$. (The 1/2 is inserted to match the usual definition of $H$ in the uniform case.)  We have chosen to measure area changes because this relates most naturally to the expansion $\theta$.\footnote{Note, however, that if we have a uniform upper bound on the extrinsic curvature $K_{ab}$ on $\Sigma$, of the form:
\begin{equation}
K_{ab} u^a u^b \le H_\text{max},\qquad \forall u^a u_a = 1,
\end{equation}
then $H(q,n^a) \le H_\text{max}$.  This bound is potentially useful if we know the values of $K_{ab}$ on $\Sigma_i$, but we don't know \emph{a priori} where $\gamma_i$ will lie on it.}

In our case, we are considering a specific lightray $\ell$ that intersects $\Sigma$.  In this case, we can take $n^a$ to be the unit vector which is proportional to the projection of the null vector $k^a$ onto $\Sigma$.  We take this as our definition of the Hubble constant $H$ at $\Sigma_i$ and $\Sigma_f$ respectively.  Importantly, this definition of the Hubble constant is independent of the null congruence that $\ell$ is a part of---it only depends on the choice of $\ell$ and $\Sigma$.

This definition of $H$ is equivalent to the following construction: let the surface $s_i \in \Sigma_i$ (or similarly $s_f \in \Sigma_f$) be defined (in a local neighborhood of $\ell \cap \Sigma_i$) by shooting out spacelike geodesics of $\Sigma_i$, in the plane of directions transverse to the lightray $\ell$.  This defines a surface $s_i$ whose extrinsic curvature satisfies $k_{ab} = 0$, and hence $k_{ab} h^{ab} = 0$.  The Hubble constant is then proportional to the \emph{expansion} of light rays shot out from $s_i$:
\begin{equation}\label{Hubble_red}
H_i =  \frac{d\lambda}{dt} \theta[s_i] =
\frac{\theta[s_i]}{(1+z)_i}
\end{equation}
where the proportionality constant depends on the redshift factor, as the Hubble constant $H$ is measured relative to proper time $dt$ rather than the affine parameter $d\lambda$.

As noted above, the parallel construction may also be used for $\Sigma_f$.  However, in this case, we can use the homogeneity and isotropy of the universe to argue that (apart from small quantum fluctuations) the geometry of $\Sigma_f$ is just Euclidean 3-space, $s_f$ is just a flat plane within that 3-geometry, and $H_f$ is equal to the usual definition of the Hubble constant.

\paragraph{Definition of Angular Scalar Excursion.}  Finally, to define the scalar excursion $\Delta \phi$ along the lightray $\ell$, we simply take the difference between the initial and final values:\footnote{If we take for granted that $\phi$ is approximately constant at the end of inflation $\Sigma_f$, we could also write $\phi(\ell \cap \Sigma_f) = \phi[\Sigma_f]$.  Similarly, if we assume that $\phi$ has remained constant since the end of inflation $\Sigma_f$, then we can equivalently compare $\phi(\ell \cap \Sigma_i)$ to its present-day value $\phi(p)$.}
\begin{equation}
|\Delta \phi| = 
|\phi(\ell \cap \Sigma_f) - \phi(\ell \cap \Sigma_i)|,
\end{equation}
(As we are currently considering a classical proof, we are neglecting quantum fluctuations, and thus we only consider classical values for $\phi$.  Equivalently, $\phi$ stands for the expectation value $\langle \phi \rangle$, in some approximately coherent state on $\Sigma_i$.)\footnote{Although our bound only cares about the values of $\phi$ along the lightray $\ell$, in cases where we don't know \textit{a priori} where $\gamma_i$ lies on $\Sigma$, it might be useful to know the set ${\cal S}$ of all values of $\phi$ that are represented on $\Sigma_i$.  In this setup, our results would imply that this set ${\cal S}$ must include \emph{at least one value} that satisfies the excursion bound relative to $\phi[\Sigma_f]$.  Understood in this way, the bound is especially stringent if we assume that ${\cal S}$ contains only a single element, i.e.~that $\phi$ is constant along $\Sigma_i$ apart from quantum fluctuations.  At the opposite extreme, the result becomes much less interesting if ${\cal S}$ includes the full range of possible values for $\phi$.  Obviously, if the full landscape of $\phi$ was \emph{already} populated on the initial slice $\Sigma_i$, then it follows that inflation \emph{wasn't needed to populate the landscape}---although it might still be needed anthropically to address the flatness problem etc.}

Alternatively, in the case of multiple scalars described by a NLSM, we use the distance measure:
\begin{equation}
|\Delta \phi| = d(\phi(\ell \cap \Sigma_f), \phi(\ell \cap \Sigma_i)).
\end{equation}

\paragraph{Proof of Bound.}
With these new angular definitions, the bound holds in the same form \eqref{eq:efoldboundlin} as before.  Specifically, if we assume the MCA, then we can find a particular light ray $\ell$ on our past lightcone such that: 
\begin{equation}\label{eq:bound2}
4\sqrt{\pi}|\Delta \phi| 
\le 
\Delta N +\log\left (\frac{H_{i}}{H_{f}}\right ),
\end{equation}
where we use the $\ell$-dependent angular definitions defined above for $\Delta N$, $H_f / H_i$, and $\Delta \phi$.  These definitions reduce that of Eq. (\ref{def}) for the isotropic metric in Eq. (\ref{eq:inflation}).

To prove this bound, we start by considering the FEB in the form \eqref{eq:CFEB}:
\begin{equation}
4\sqrt{\pi}|\Delta \phi|
\le
\log\left ( \frac{\theta[\gamma_i]}{\theta[\gamma_f]}\right ),
\end{equation}
where we have removed the absolute value signs around the RHS because (as will be shown below) $\theta[\gamma_i] \ge \theta[\gamma_f] > 0$.

On the final slice $\Sigma_f$, we have:
\begin{equation}
\theta[\gamma_f] \approx \theta[s_f],
\end{equation}
where this relation holds to a very high degree of accuracy because, as stated above, $\gamma_f$ is a very large sphere in units of $1/H_f$, so its inverse radius is approximately zero, and its $\tr\,k$ is thus very close to zero.  It follows from \eqref{Hubble_red} that
\begin{equation}
\theta[\gamma_f] \approx 2H_f (1+z)_f > 0.
\end{equation}
By the focusing of light rays due to the NEC, it then follows that $\theta$ must be positive and greater on $\partial P$ at all earlier times.

On the initial timeslice $\Sigma_i$, we have 
\begin{equation}
\theta[s_i] \ge \theta[\kappa_i] \ge \theta[\gamma_i] \ge 0.
\end{equation}
Here, the first inequality holds by construction, because $s_i$ has $\tr\,k = 0$, while $\kappa_i$ has $\tr\, k > 0$ (for the outward normal $n^a$, which points in the opposite direction to $k^a$).  The second inequality holds by virtue of the MCA, because $\gamma_i$ is marginally bounded by $\kappa_i$, meaning that $\gamma_i$ (unless it coincides with $\kappa_i$ in a neighborhood of the point $q$) can only deviate from $\kappa_i$ in an inward direction, not an outward direction (see the discussion of this geometric fact in \cite{Wall:2010jtc,Wall:2012uf,Engelhardt:2013tra}).  And the third inequality has already been stated above.

Again invoking \eqref{Hubble_red}, it follows that
\begin{equation}
\theta[\gamma_i] \le 2H_i (1+z)_i.
\end{equation}
Hence, the ratio satisfies
\begin{equation}
\frac{\theta[\gamma_i]}{\theta[\gamma_f]}
\le
\frac{H_i}{H_f} \frac{(1+z)_i}{(1+z)_f}
\end{equation}
which proves the bound \eqref{eq:bound2} as stated above.

\subsection{Applications to phenomenology}
\label{sec:apppheno}
As an example of how this result might affect phenomenology, let us suppose we are trying to use inflation to solve the fine-tuning problems of theoretical particle physics, by postulating a large landscape of $n_v$ possible vacua.\footnote{Here, we leave aside the important philosophical question about whether a multiverse would count as a good explanation for fine-tuning (see e.g.~\cite{manson2022cosmic,isaacs2022multiple} and citations therein).  We confine our analysis to the physics question of whether there is enough time to reach the requisite number of vacua.}  To illustrate the point in the starkest possible terms, let us suppose that $\ell_\text{string} \sim \ell_\text{SUSY} \sim \ell_\text{KK} \sim \ell_\text{planck}$ (assuming that, respectively, string theory, supersymmetry, and/or Kaluza-Klein dimensions exist), and that at energy scales below $\ell_\text{planck}^{-1}$, physics is well-described by a traditional renormalizable Wilsonian QFT paradigm in $D = 4$ spacetime dimensions.  (In other words, we take away all other resources that would be helpful for reducing the amount of fine-tuning, and try to solve it with a landscape alone.)  Between the cosmological constant (tuned to $\sim\!10^{-120}$) and the Higgs mass squared (tuned to $\sim\!10^{-30}$), we would need $n_v \gtrsim 10^{150}$ distinct vacua to expect to find at least one vacuum that is as fine-tuned as our universe.\footnote{If we instead postulate low energy supersymmetry ($\ell_\text{SUSY} \sim 100 \text{ TeV}$), the amount of fine-tuning required drops to around $10^{-60}$, and it is also permissible for the scalar potentials to be less generic.  But this requires restricting to supersymmetric regions of moduli space, and it is not totally clear whether such vacua are more numerous in the string landscape than vacua with Planck scale (or no) supersymmetry \cite{Susskind:2004uv}, even though the latter require more fine-tuning.}

In order to apply our results, we also ignore discrete parameters such as fluxes, and restrict our attention to landscapes which involve varying $n_s$ different scalar fields $\phi^A$ alone, with some potential $V(\phi^A)$ where a vacuum is defined as a critical point of the potential ($V' = 0$) which is stable ($V''$ is a positive-definite matrix), and the cosmological constant is the value of $V_0$ at the critical point. If we zoom in on any such vacuum, in any region of moduli space shaped like a ball $B$ (in terms of excursion length measured by the metric $d(X_1, X_2)$) with a radius $r \ll 1$ (i.e., small in Planck units), we can then neglect irrelevant couplings.  (As this is true for any $r \ll 1$, for purposes of estimating the number of vacua we can consider a ball $B$ with $r \sim 1$, using a slightly more generous definition of $\sim$.)  The low-energy QFT potential takes the renormalizable form:
\begin{equation}\label{V}
V = V_0 
+ \lambda^{(2)}_{AB} \phi^A \phi^B
+ \lambda^{(3)}_{ABC} \phi^A \phi^B \phi^C
+ \lambda^{(4)}_{ABCD} \phi^A \phi^B \phi^C \phi^D+\ldots,
\end{equation}
where we ignore $\phi^5$ and higher terms as they are irrelevant in $D = 4$.  In a generic situation with no symmetry, we would expect $\lambda^{(4)}_{AB}$ to take a generic positive quartic form.\footnote{There might be ways of cleverly engineering additional vacua using e.g.~non-polynomial terms (from radiative corrections) in effective quantum potentials, but we assume this is unlikely to produce an enormously greater number of vacua.}   The equation $V' = 0$ is thus a system of $n_s$ cubic equations.  For a generic quartic potential, B\'{e}zout's theorem\footnote{Bézout's theorem asserts that a system of $n$ polynomial equations in $n$ variables has exactly 
$\prod_{i=1}^n \deg(f_i)$ solutions (counted with multiplicity, and including complex solutions as well as projective solutions ``at infinity''), except in certain nongeneric cases, where the number of solutions is instead infinite.} implies that there are exactly $3^{n_{s}}$ distinct critical points, which might, however, be at either real or complex values of the $\phi^A$ coordinates.  In our case we are interested only in the critical points which are i) real, ii) stable, and iii) within the regime of validity of the potential \eqref{V}, so we get a (not always tight) upper bound of $3^{n_{s}}$ stable vacua within the ball $B$, i.e.~that are $\ll 1$ Planck-excursion from the original vacuum.\footnote{A simple example of a potential with $3^{n_s}$ real critical points is $V = \sum_A \frac{1}{4}(\phi^A)^4 - \frac{1}{2}(\phi^A)^2$.  In this case, only $2^{n_s}$ of them are stable.  Surprisingly, this is not the actual upper bound!  In the case $n_s = 2$, \cite{durfee1993counting} proved that there cannot be more than five stable critical points, while the functions described in \cite{MOexample} and \cite{MSEexample} show that this can, in fact, be achieved.}  Of course, there may also be many other planck-sized regions of moduli space that contain \emph{no} vacua.

If we further assume (admittedly a highly contestable assumption) that the scalar moduli space carries a flat metric, then enlarging the radius $r$ of our search ball increases the accessible volume as $r^{n_s}$.  (The unit sphere volume factor $\text{Vol}(S_{n_s - 1})/n_s$ is the same for both spheres and thus cancels in the ratio.)  Consequently, ensuring the existence of at least one viable vacuum would require a minimum field excursion of order
\begin{equation}
\Delta \phi \gtrsim \frac{10^{150/n_s}}{3} .
\end{equation}
In a nearly classical slow-roll inflationary scenario, this translates into a required number of e-folds
\begin{equation}
\Delta N \gtrsim \frac{4\sqrt{\pi}}{3}10^{150/n_s},
\end{equation}
in order to expect at least one viable vacuum within the explored range. Unless the number of scalar fields satisfies $n_s \gtrsim 75$, this exceeds the $30$ to $60$ e-folds usually invoked to explain observation. Even so, these estimates are generous understatements: typical Planck-scale balls in moduli space are unlikely to contain any critical points at all. As we currently lack a reliable method for estimating the typical number, we do not attempt a sharper refinement here.

It might, however, be possible to have many fewer e-folds if $n_s \ge 2$, and the moduli space of the scalar fields is negatively curved (hyperbolic), as then the amount of available volume grows exponentially in 
$r$.  By a similar renormalization argument to the above, we expect the curvature length $L$ of the non-linear sigma model to be at most Planckian.  If the moduli space has the geometry of $n_s$-dimensional, maximally-symmetric hyperbolic space, the ratio of the volume of a ball of radius, $r$, to the unit ball scales like:
\begin{equation}
\frac{\text{Vol}(H_r)}{\text{Vol}(\text{unit }B)} = n_s \int^r_0 \sinh^{n_s - 1}(r'/L)\,dr' \approx 
\frac{n_s}{2^{n_s-1}(n_s-1)} \exp\!\left((n_s - 1) r / L\right)
\end{equation}
using the large excursion approximation $r \gg L$.  Neglecting the log of the $n_s$-dependent prefactors (including the factor of $3^{n_s}$ for the maximum number of nearby vacua), we need an excursion of at least
\begin{equation}
(n_s - 1) \frac{\Delta \phi}{L} \gtrsim 150 \log(10) \approx 345,
\end{equation}
and hence the number of e-folds required by the bound
\begin{equation}
\Delta N \gtrsim \left ( \frac{(4\sqrt{\pi})}{n_{s}-1}150\log(10)\right )L \approx (2.5\times 10^{3})\frac{L}{n_{s}-1}.
\end{equation}
Unless $n_s \gg 1$, this bound is likely to require more e-folds than the minimum number required by observation.  Again, this bound generously assumes $O(1)$ vacua are present in each Planck-sized ball of moduli space, so the actual lower bound is likely to be more stringent.

The central point is that if an inflationary multiverse truly accounts for fine-tuning, then (unless there are a rather large number of scalar fields) the inflationary epoch must go back much earlier than the current observational lower bound. This expectation contrasts sharply with the Hartle–Hawking wavefunction proposal \cite{Hartle:1983ai}.  Although the proposal appears to give reasonable predictions for the high-energy modes of the quantum fluctuations \cite{Mukhanov:1981xt,Starobinsky:1982ee,Guth:1982ec,Hawking:1982cz,Bardeen:1983qw}, it favors an inflationary period with far fewer e-folds than even observational bounds permit \cite{Maldacena:2024uhs}. In fact, if we do not condition on inflation happening at all, the Hartle-Hawking proposal predicts that an even more dominant contribution to the saddle should be a big empty de Sitter universe with the present-day value of $\Lambda$, dramatically contradicting observation.

These issues could potentially be remedied by considering another variant of the no-boundary proposal, for example the Vilenkin tunnelling proposal\footnote{This model was originally proposed in \cite{vilenkin1982creation}, for more recent discussion see Vilenkin and Yamada \cite{Vilenkin:2018dch,Vilenkin:2018oja} and citations therein.} in which the universe tunnels from zero size ($a = 0$) along a Lorentzian contour.  Such models typically make the opposite prediction from the Hartle-Hawking model, that the universe should start out as small as possible, and thus at a high value of the inflationary potential.  It was argued in \cite{Feldbrugge:2017kzv} that this model gives an unnormalizable (inverse Gaussian) prediction for the quantum fluctuations, but Vilenkin and Yamada claim that this issue can be resolved by adding an appropriate boundary term to the action; see also \cite{DiTucci:2019dji} for a similar idea.

Thus, if the initial condition was not Hartle-Hawking, it is more plausible that the universe started at a point in the landscape from which a large amount of inflation was possible. It is also possible for inflation never to end (see e.g.~\cite{Guth:2007ng} for a review of this ``eternal inflation''), but as such scenarios typically involve quantum fluctuations, the above conclusions will not necessarily be applicable unless the FEB can be generalized to quantum situations.  This will be the subject of the next section.

\section{Quantum effects}\label{quantumeff}

We now comment on a possible quantum generalization. As stated above Eq. (\ref{eq:einsteineqR}), we assumed the NEC in this derivation. However, the NEC can be violated by quantum effects \cite{Epstein:1965zza} such as Hawking radiation. We need to consider a quantum generalization of the NEC, which will involve accounting for entropy variation. We will keep Newton's constant, $G_{\textrm{N}}$, explicit in this section because its renormalization will play an important role. 

\begin{figure}[t]
\centering
\includegraphics[width=8cm]{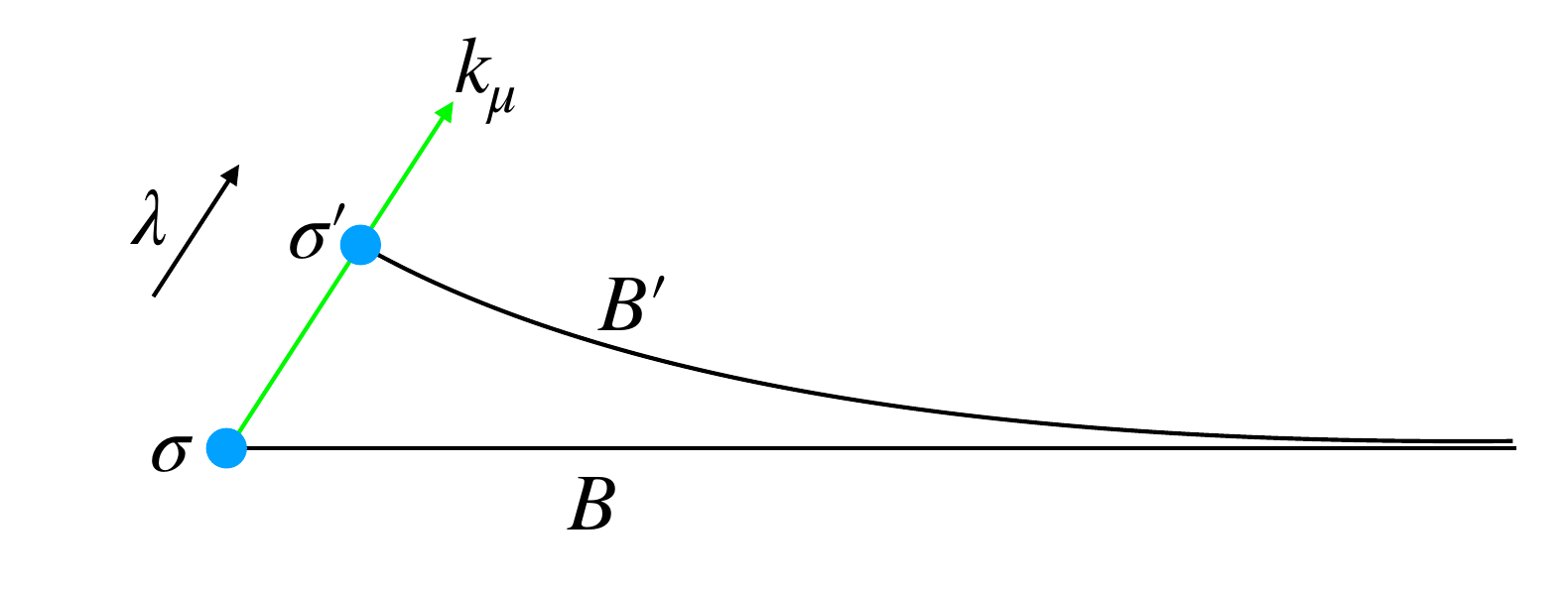}
\caption{An illustration of the spacelike surfaces $B$ and $B'$ bounded by spacelike surfaces $\sigma$ and $\sigma'$ related by evolution along the $k_{\mu}$ null vector.}
\label{fig1}
\end{figure}

\begin{figure}
\centering
\includegraphics[width=6cm]{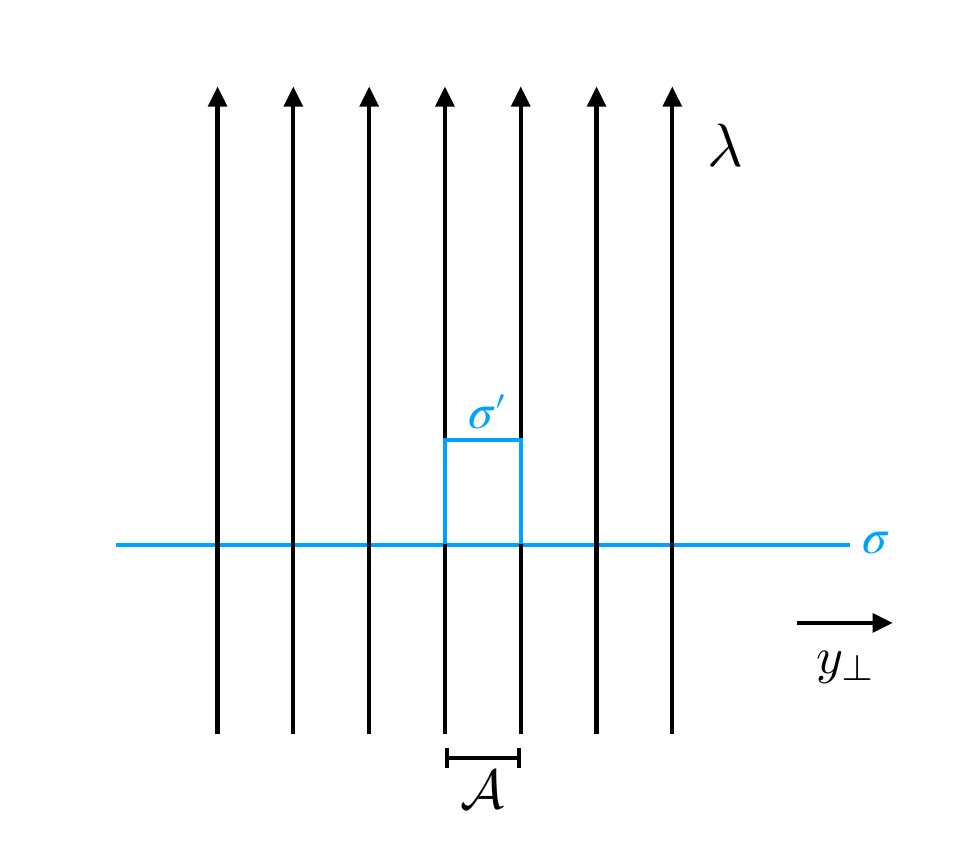}
\caption{An illustration of how the derivative measures how the entropy changes relative to an infinitesimal change in $\sigma$ localized in an area $\mathcal{A}$ around $y_{1,\perp}$.}
\label{fig2}
\end{figure}

We consider the entropy associated with codimension-one spacelike hypersurfaces $B$, each bounded by a family of codimension-two spacelike surfaces defined by the embedding function $ \sigma_{\lambda}(y_{\perp}) $; see Fig.~\ref{fig1}. Here, $\lambda$ is an affine parameter labeling the spacelike slices, while $ y_{\perp} $ denotes coordinates on a fixed-$\lambda$ slice. We consider the generalized entropy
\begin{equation}\label{eq:Sgendef}
S_\text{gen}=\frac{A}{4G_{\textrm{N}}}+S_\text{mat}
\end{equation}
associated with a given $\sigma_{\lambda}(y_{\perp})$. $S_\text{gen}$ generalizes the notion of area when quantum effects are important. Rather than studying how the entropy and area evolve as a function of $ \lambda$, we treat the area and entropy as functionals of $ \sigma_{\lambda}(y_{\perp}) $ and consider functional derivatives localized in a small area $\mathcal{A}$ around $y_{1,\perp}$; see Fig.~\ref{fig2}. Using the same notation as Ref. \cite{Bousso:2015mna}, we use the shorthand 
\begin{equation}
S'=k^{\mu} \frac{\delta S}{\delta \sigma_{\lambda}^{\mu}(y_{\perp})}|_{y_{1,\perp}} \ .
\end{equation}
Since entropy can be highly non-local, $S'$ can depend on the choice of spacelike slicing far away from $y_{1,\perp}$. We therfore parameterize how $S_\text{gen}$ locally changes along $k_{\mu}$ using the quantum expansion parameter:
\begin{equation}\label{eq:quantumexp}
\Theta= \lim_{\mathcal{A}\rightarrow 0}\frac{4G_{\textrm{N}}}{\mathcal{A}}S_\text{gen}'\ \bigg |_{y_{1,\perp}} \ .
\end{equation}
The quantum expansion parameter
gives the rate that $S_\text{gen}$ increases per unit area along the light ray $k_{\mu}$ and reduces to the classical expansion parameter when the matter entropy is null. 

The original FEB came from assuming the focusing condition 
\begin{equation}\label{focusingcon}
\frac{d\theta}{d\lambda}\leq \frac{-1}{D-2}\theta^{2}\ ,
\end{equation}
and then adding a contribution from a coherently varying field along the light ray. Therefore, the natural quantum generalization comes from assuming a strengthened quantum focusing conjecture (SQFC),\footnote{Aron Wall: This bound was originally proposed to me by Andrew Rolph in a private communication, April 2022.  Later, my coauthor, Aidan Herderschee, thought of the SQFC independently.  Due to a memory lapse, I did not credit Rolph in the original preprint version of this paper.  I apologize for the oversight.}
\begin{equation}\label{strenqfc}
\textrm{SQFC}: \quad \frac{d\Theta}{d\lambda}\leq \frac{-1}{D-2}\Theta^{2} \ ,
\end{equation}
motivated by the idea that quantum focusing inequality should have a definite-sign quadratic term just as the classical Raychaudhuri Eq. does. Then we consider the contribution of a coherently varying field along the light ray. We refer to Eq. (\ref{strenqfc}) as the SQFC, because the original quantum focusing conjecture (QFC) only imposed that the left-hand side was negative \cite{Bousso:2015mna}. 

Importantly, we note that the SQFC is sensitive to the renormalization of Newton's constant. In terms of $S_{\text{gen}}$, the SQFC reduces to
\begin{equation}\label{sqfcexplicitz}
\partial_{\lambda}\left (\frac{S_{\text{gen}}'}{\mathcal{A}} \right )\leq - 4G_{\textrm{N}}\left (\frac{S_{\text{gen}}'}{\mathcal{A}}\right )^{2} \ .
\end{equation}
In the original QFC, the entropy suffers from UV divergences, but these can be absorbed into the renormalization of Newton's constant, so $\Theta$ is well-defined \cite{Susskind:1994sm,Bousso:2015mna}. More concretely, the leading order area-law renormalization leads to the replacements 
\begin{equation}\label{eq:reshift}
S_{\textrm{mat}}\rightarrow S_{\textrm{mat}}+ b\mathcal{A}, \quad \frac{1}{G_{\textrm{N}}}\rightarrow \frac{1}{G_{\textrm{N}}}-4b
\end{equation}
for some constant $b$, where $G_{\textrm{N}}$ is Newton's constant. The shift in Eq. (\ref{eq:reshift}) leaves $S_{\textrm{gen}}$ (\ref{eq:Sgendef}) invariant. In contrast, a shift in Newton's constant due to renormalization changes the relative sizes of the two terms in Eq. (\ref{sqfcexplicitz}) due to the explicit factor of $G_{\textrm{N}}$.  The meaning of the bound thus requires picking some choice of $G_{\textrm{N}}$, which is a nontrivial decision if $G_{\textrm{N}}$ is subject to RG flow.
As a conservative choice, we propose that whenever the SQFC is applied to an EFT calculation valid between some minimum energy scale $E_\mathrm{min}$ and maximum energy scale $E_\mathrm{max}$, that (i) an effective value of $G_N(E)$ should be defined at each energy $E_\text{min} < E < E_\text{max}$ using the area-law formula, and that (ii) the SQFC holds for 
whichever of these values of $G_N(E)$ is the smallest (i.e.~the least stringent choice in \eqref{sqfcexplicitz}).

For example, if we consider a CFT in 2+1 dimensions, a disk of radius $r$ in the vacuum will have entropy
\begin{equation}\label{disk}
S_\textrm{mat} = c_0 + c_1 r,
\end{equation}
and an ``area law'' scheme for $G_\text{N}$ will set $c_1 = 0$, so that the area-law part of the generalized entropy is entirely contained in the $A/4G_\text{N}$ term.  A more complicated story occurs if the matter sector flows from a UV CFT to an IR CFT.  In this case, Eq. \eqref{disk} only holds asymptotically in either the $r \to 0$ limit (the UV CFT) or the $r \to \infty$ limit (the IR CFT).  It has been proven that $S''(r) < 0$ \cite{Casini:2012ei}, so it is not possible to have $c_1 = 0$ in both the UV and the IR.  In fact, $G_\text{UV} < G_\text{IR}$ (i.e.~in this area-law scheme gravity is anti-screening).  In the case of the future lightcone of a point $p$ in the 2+1 Minkowski vacuum, we have checked that the SQFC holds when defined in terms of $G_\text{UV}$, by setting $c_1 = 0$ in the UV.  (This implies that $S'(r) \le 0$ for all $r > 0$, and thus ensures that $\theta$ and $S'$ have the opposite signs.)

In higher dimensions, it might be necessary to also think carefully about subleading divergences, beyond the area law, which are associated with higher curvature terms in the effective gravitational action (cf.~\cite{Bousso:2015mna} and citations therein).

The strengthened QFC implies some desirable properties about spacetime, such as quantum generalizations of the Penrose singularity theorem \cite{Penrose:1964wq,Hawking:1965mf,Hawking:1973uf}.  It was previously argued by one of us that the Generalized Second Law (GSL), which is implied by the QFC, can be used to prove the incompleteness of null geodesics from the existence of a ``quantum trapped surface'' $\cal T$ with $\Theta \le 0$ everywhere \cite{Wall:2010jtc}.\footnote{Together with the auxiliary assumptions of global hyperbolicity and spatial noncompactness, which are also used in the classical Penrose theorem.}  The SQFC would allow one to strengthen this result by deriving that the termination should occur before an affine time $\Delta \lambda \le -(D-2)/\Theta[\mathcal{T}]$, analogously to the classical result.

We further note that the SQFC is similar to a bound proposed by Ben-Dayan \cite{Ben-Dayan:2023inz}.  Although the details of Ben-Dayan's inequality vary from our own, it is similarly obtained by adding to the QFC a term which vanishes when $\Theta = 0$.  All such bounds will imply the restricted QNEC, that $d\Theta/d\lambda \le 0$ whenever $\Theta = 0$, which has been proven in the context of the brane-worlds scenario \cite{Shahbazi-Moghaddam:2022hbw}.

We now turn back to bounding field excursions. The core intuition behind the original FEB’s derivation from the NEC is that, if the other stress–energy contributions violated the NEC, one could simply eliminate the field excursion and reveal a NEC violation. We adopt a similar line of reasoning here. We assume that the expectation values of the scalar field and the area operator obey Einstein’s equations. To derive a quantum FEB (QFEB), we consider an original state where the scalar expectation value $\langle \phi \rangle$ changes along the null ray. We then act on this state with an operator that removes the coherent variation of $\phi$ within a local region and apply Eq. (\ref{strenqfc}) to this state.  Thus, in the new state $\langle \phi \rangle = \textrm{constant}$.  (As the local region can be small, the required shift in $\Delta \langle \phi \rangle$ can be made arbitrarily small, and thus should not significantly affect the behavior of the matter quantum state through direct couplings.)  In order for Eq. (\ref{strenqfc}) to hold for the new state, the following bound must hold on the original state with coherent field variation:
\begin{equation}\label{strenqfc2}
\left | \frac{d\Theta}{d\lambda}\right |\geq  \frac{1}{D-2}\Theta^{2}+8\pi \left ( \frac{d\phi}{d\lambda}\right )^{2}\ ,
\end{equation} 
given that the operator removing the local coherent variation in $\phi$ does not change the entropy $S_\text{mat}$, as this is a purely classical shift of the quantum fields. We can then apply the same argument as in Section \ref{febsec} to find that 
\begin{equation}\label{QFEB}
\textrm{QFEB}: \quad \left |\log\left ( \frac{\Theta_{2}}{\Theta_{1}}\right ) \right |\geq 4\sqrt{\frac{2\pi}{D-2}} d(X_{2},X_{1}) \ .
\end{equation}
where the implicit Newton's constant is the minimal value permitted within the relevant EFT regime. The kinetic term of the scalar field is also subject to renormalization, affecting the definition of $d(X_2,X_1)$.  This renormalization should also be considered carefully, perhaps by a similar method to what we proposed for $G_\text{N}$, using the most conservative effective value of the parameter in the RG flow.

A proof of the SQFC, and therefore the QFEB, is beyond the scope of this paper. Even the validity of the original QFC is still unproven in quantum situations that go beyond the $G_{\text{N}} \to 0$ (QNEC) limit.\footnote{In fact, Ref.~\cite{Shahbazi-Moghaddam:2022hbw} suggests that the QFC, and hence the SQFC, may be violated for certain large-$N$ CFTs coupled to gravity for specific choices of spacelike slicing.} Nevertheless, it is noteworthy that the QFEB holds in spacetimes that do satisfy the SQFC.

Finally, we comment on the implications of quantum effects for phenomenology. While the QFEB is interesting, it appears to have limited utility in concrete phenomenological applications. Instead, the central question is under what circumstances quantum corrections could invalidate the arguments of Section \ref{sec:apppheno}. For such an invalidation to occur, a model must satisfy two conditions simultaneously: it must exhibit genuinely “fast” field excursions and feature quantum effects that dominate over the scalar field’s kinetic term in the Raychaudhuri equation. By ``fast'', we mean that the FEB is close to saturation, or at least some order one fraction away from saturation. If the FEB is far from being saturated, violating it in a spacetime with a positive cosmological constant would require extraordinarily large quantum corrections, strong enough to generate focusing effects that rival the redshift from cosmic expansion, which seems highly implausible. Conversely, suppose quantum effects do not dominate over the kinetic term. In that case, one still generally obtains an approximate linear bound on the field excursion in terms of the change in e-folds, even if the original FEB is perturbatively violated.\footnote{Even when the quantum contribution is comparable to the kinetic term, a linear bound can still persist, though with a modified numerical coefficient.} Although we cannot strictly exclude the possibility of models in which both criteria are met, such scenarios would appear to be highly exotic. Notably, the first condition is not obeyed in models obeying the slow-roll approximation,\footnote{Although certain cosmic microwave background measurements place strong upper bounds on the slow-roll parameter \cite{Planck:2018jri}, these constraints apply only within specific models and to the relatively recent cosmological history, roughly the last 60 e-folds of inflation.} which encompasses most scenarios of eternal inflation \cite{1983veu..conf..251S,Vilenkin:1983xq,Guth:2007ng}. Of course, by definition, eternal inflation involves arbitrarily large values of $\Delta N$, but inflation is still past-incomplete \cite{Borde:2001nh}, and it is still interesting to bound the amount of excursion that can take place in some finite number of e-folds. 

\section{Discussion}\label{disc}

In this paper, we derived a novel bound on field displacement along a null geodesic in terms of the expansion parameter, $\theta$. We applied our bound in the context of a naked singularity and large-field inflation. We also suggested a quantum generalization of the bound based on a strengthened QFC. This paper should be considered an initial foray into bounding field excursions in gravitating regions using modern proof techniques. An important feature of the FEB is its dependence on the full trajectory of the light ray, not merely on data at its endpoints. This arises from its explicit dependence on the affine parameter. 

In the context of inflation, our result bears a close resemblance to the Lyth bound \cite{Lyth:1996im}, which provides a lower limit on the scalar field excursion in terms of the tensor-to-scalar ratio $r$ and the number of e-folds,  
\begin{equation}
\Delta \phi \geq \mathcal{O}(1)\,\sqrt{r}\,\Delta N \ .
\end{equation}
The Lyth bound is important because it relates observable primordial gravitational waves to super-Planckian field excursions, offering a direct probe of high-scale inflation through measurements of the cosmic microwave background. By contrast, the FEB places an \emph{upper} limit, not a lower one, on the total displacement of the inflaton and does so independently of the shape of the potential. This distinction makes the FEB particularly valuable for constraining the dynamics of the inflaton in the very early universe, well before the last \( 60 \) e-folds, where we (currently) lack direct observational access.

We emphasize that our excursion bound follows directly from Einstein’s equations, and therefore cannot exclude any cosmological model consistent with them. At the same time, the bound is a powerful tool for ruling out classes of models with particular features. For instance, it yields a no-go theorem against the construction of short-duration large-field inflation models without large violations of the NEC. 

When formulating the quantum generalization of the FEB, we were required to assume the strengthened QFC. The strengthened QFC is a novel and intriguing condition on gravitational theories, deserving exploration in its own right. In particular, it would be valuable to investigate explicit examples in higher-dimensional settings to test its validity. Moreover, just as the original QFC led to the QNEC \cite{Bousso:2015wca,Wall:2017blw,Balakrishnan:2017bjg,Ceyhan:2018zfg} in the non-gravitational limit, it is natural to ask whether the SQFC could imply new, nontrivial constraints on non-gravitational quantum field theories. 

One might wonder whether invoking the SQFC is actually necessary to derive a bound on the field excursion in terms of the entropy, expansion parameter, and/or area. For instance, one could have hoped that the QNEC would suffice to establish a bound on $d(X_{2},X_{1})$. Unfortunately, while the QNEC is in principle applicable in the semiclassical regime where gravity is treated as an effective field theory, it does not yield a local bound on $d(X_{1},X_{2})$ that is independent of the detailed behavior of the entropy along the null ray.

In addition to quantum effects, one should also consider higher-derivative corrections. In many phenomenologically plausible models of quantum gravity, such corrections are suppressed by the Planck scale and are therefore expected to be small. Although rigid source terms like $J(x)\phi(x)$ and non-derivative higher-dimension operators do not contribute to Eq.~\eqref{eq:scalarsource}, higher-derivative terms, such as
\begin{equation}\label{eq:highderterm}
S \supset \int d^{d}x \ \sqrt{-g}\ (\partial_{\mu} \phi \partial^{\mu} \phi)^2
\end{equation}
can contribute. In the presence of such terms, it is unclear whether a version of the FEB continues to hold. These terms are particularly sensitive to variations of the field in directions transverse to the null geodesic. Interestingly, the Wilson coefficients of such operators are constrained by positivity bounds \cite{Adams:2006sv}.

One might ask what form a bound based solely on the area would take. Unlike the FEB, a bound expressed purely in terms of the area at the endpoints of a congruence would have the appealing feature of depending only on data defined on the initial and final spacelike slices bounding the light ray. A natural candidate, consistent with known examples~\cite{Dolan:2017vmn, Draper:2019zbb, Draper:2019utz, Delgado:2022dkz}, is  
\begin{equation}\label{eq:areab}
\left| \log\left(\frac{\mathcal{A}_{2}}{\mathcal{A}_1} \right) \right| \geq \alpha\, d(X_1, X_2) \ ,
\end{equation}
for some constant $\alpha$.\footnote{Eq.~\eqref{eq:areab} is a classical inequality and can be violated in spacetimes that do not satisfy the null energy condition (NEC).} This expression can be viewed as a covariant generalization of the proposal in Ref.~\cite{Draper:2019utz} that field excursions in a localized region of spacetime should be bounded by the volume of the spacetime region.\footnote{A bound in terms of spatial volume is excluded for the same reasons that an entropy bound in terms of volume is inconsistent~\cite{Bousso:1999xy,Bousso:2002ju}.} However, Eq.~\eqref{eq:areab} fails near naked timelike singularities, such as the example discussed in Section~\ref{sec:nakedsingular}, at a finite affine distance from the singularity.\footnote{We thank Craig Clark for emphasizing this point.} Thus, any derivation of Eq.~\eqref{eq:areab} would necessarily rely on, or amount to a proof of, some version of cosmic censorship~\cite{Penrose:1969pc}, a notoriously subtle and unresolved issue.\footnote{See Ref.~\cite{Engelhardt:2024hpe} for a recent proposal.} For example, a possible counterexample to the area bound (\ref{eq:areab}) could be constructed by taking a spacetime with a naked timelike singularity, excising an arbitrarily small region around the singularity, and replacing it with a nonsingular spacetime. However, we show how energy conditions can exclude a large family of these counterexamples in Appendix~\ref{sec:cutoutnsing}. 

It would be interesting to explore whether the bounds derived in this paper have any implications for the swampland program. Despite their relation to certain swampland conjectures, our bounds are conceptually distinct. For example, the infinite distance conjecture \cite{Ooguri:2006in} concerns the behavior of effective field theories near points at infinite distance in the vacuum manifold, whereas the FEB places local constraints on the ability to probe large distances in field space. Similarly, although the trans-Planckian censorship conjecture \cite{Bedroya:2019snp} bears a superficial resemblance to the FEB, it pertains to the suppression of sub-Planckian quantum fluctuations becoming classical, rather than to classical trans-Planckian field excursions in the vacuum manifold.

\subsection*{Acknowledgments}

We thank Carolina Figueiredo and Qianshu Lu for their collaboration in the early stages of this work. We thank Chris Akers, Nima Arkani-Hamed, Scott Collier, Åsmund Folkestad, Sergio Hernández-Cuenca, Patrick Jefferson, Zohar Kormagodski, Jonah Kudler-Flam, Juan Maldacena, Andrew Rolph, Jonathan Source, Douglas Stanford, Evita Verheijden, Edward Witten, and Bowen Zhao for helpful discussions.  AH is grateful to the Simons Foundation as well as the Edward and Kiyomi Baird Founders’ Circle Member Recognition for their support.  Aron Wall was supported by the AFOSR grant FA9550-19-1-0260 “Tensor Networks and Holographic Spacetime”, and the STFC grant ST/X000664/1 “Quantum Fields, Quantum Gravity and Quantum Particles”.  He was also supported by NSF grant PHY-2207584 during his sabbatical at the IAS.

\appendix

\section{Restrictions on cutting out naked singularities}\label{sec:cutoutnsing}

There is some reason to hope for an area bound on field excursions under the assumption of cosmic censorship. However, one challenge in using cosmic censorship to prove Eq. (\ref{eq:areab}) is that one can naively excise a region near a naked singularity and still find that Eq. (\ref{eq:areab}) is violated. Remarkably, as first suggested in Ref. \cite{Nicolis:2008wh}, energy conditions imply that a finite region surrounding the singularity must also be removed when excising the singularity itself. The hope is that this minimal cut-out region captures all the loci where Eq. (\ref{eq:areab}) fails. If so, this would suggest that a proof of Eq. (\ref{eq:areab}) might be possible based on cosmic censorship and local energy conditions. While we cannot rigorously establish this, we provide supporting evidence when the parameter $\beta$ is sufficiently small.

To see why a minimal region must be excised along with the naked singularity, we appeal to the concept of positive quasi-local mass. Suppose we excise a spherical region of radius $r$ centered on the singularity and replace it with another spacetime patch. The quasi-local mass of the inserted region can often be computed from boundary data alone, by Stokes' theorem, and is therefore independent of the interior geometry. However, defining a positive-definite quasi-local mass in general spacetimes is notoriously subtle. To sidestep this difficulty, we assume the spacetime is stationary and radially symmetric, with non-negative energy density. Under these assumptions, the Hawking quasi-local mass \cite{Misner:1964je,Hernandez:1966zia,Hawking:1968qt} enclosed by a sphere centered at the origin is guaranteed to be positive \cite{Christodoulou1988SomeRO}. If the Hawking mass becomes negative sufficiently close to a naked timelike singularity, it implies that any sewn-in spacetime preserving positive energy conditions must exclude a neighborhood of the singularity.

We define the quasi-local energy using the Hawking quasi-mass instead of the quasi-local mass in Ref. \cite{Nicolis:2008wh}, which is not positive-definite away from the weak field regime.  Given the metric 
\begin{equation}
ds^{2}=-A(R)dt^{2}+B(R)dR^{2}+R^{2}d\Omega^{2}
\end{equation}
the Hawking quasi-local mass is \cite{Faraoni:2020mdf} 
\begin{equation}\label{eq:mrtdef}
m(R)=\frac{R}{2}\left (1-\frac{1}{B(R)} \right )\ .
\end{equation}
Again, Eq. (\ref{eq:mrtdef}) is positive definite assuming that the matter-energy density is positive definite \cite{Christodoulou1988SomeRO}. In fact, Ref. \cite{Christodoulou1988SomeRO} shows that the Hawking quasi-local mass obeys the stronger bound, 
\begin{equation}\label{strongerpos}
m(R)>\frac{1}{12\pi}\sqrt{\frac{\textrm{Area}_{S}}{16 \pi}}\int_{S} R_{M}
\end{equation}
where $M$ is the spacelike slice on which we are computing the mass and $S$ is the codimension two spacelike surface that encloses the region of interest. 

We now consider the metric in Eq. (\ref{eq:singmetrci}). From Eq. (\ref{eq:mrtdef}), the quasi-local mass contained within a sphere of radius $r$ is
\begin{equation}\label{eq:solutionquasilocmass}
m(r)=-\frac{1}{2} m r^{\frac{\beta -1}{2}} (r-2 m)^{\frac{1}{2} (-\beta -1)} \left((\beta +1)^2 m-2 \beta  r\right) \ .
\end{equation}
A similar calculation shows that the right-hand side of the bound in Eq. (\ref{strongerpos}) is 
\begin{equation}
m(r)>-\frac{1}{3} \left(\beta ^2-1\right) m^2 r^{\frac{\beta }{2}-\frac{1}{2}} (r-2 m)^{\frac{1}{2} (-\beta -1)}
\end{equation}
The bound on the mass (\ref{strongerpos}) is violated when 
\begin{equation}\label{eq:conditionnegativemass}
\textrm{Mass Positivity Violation}:\quad r\leq \frac{1}{6} \left(\beta +\frac{5}{\beta }+6\right) m
\end{equation}
Violation of mass positivity for sufficiently small $r$ comes from how the naked singularity contributes negative energy and violates energy conditions. The argument in the previous paragraph is therefore applicable, and we have demonstrated how removing a naked singularity also involves removing a finite region around the singularity. 

While we have argued that a finite region around the singularity must be cut out, we find this particular argument is insufficient to save Eq. (\ref{eq:areab}) for generic $\beta$. Along an infinitesimal region of the null geodesic, Eq. (\ref{eq:areab}) becomes the condition 
\begin{equation}
\frac{1}{\sqrt{h_{\sigma}}}\frac{d\sqrt{h_{\sigma}}}{dr}\geq \alpha \frac{d\phi}{dr} \ .
\end{equation}
Using that 
\begin{equation}
\sqrt{h_{\sigma}}\propto r^{2} f^{1-\beta}
\end{equation}
we find that Eq. (\ref{eq:areab}) is violated for 
\begin{equation}\label{eq:areaboubd}
\textrm{Violate Area Bound}: \quad r\leq m \left(\frac{\alpha}{4\sqrt{\pi}}\sqrt{1-\beta ^2}+\beta +1\right)
\end{equation}
Comparing Eq.~(\ref{eq:areaboubd}) to Eq.~(\ref{eq:conditionnegativemass}), we see that the area bound cannot be violated by spacetimes constructed by excising the singularity only for
\begin{equation}\label{eq:badbeta}
1>\frac{10 \sqrt{\pi }}{\sqrt{100 \pi + 9 \alpha ^2}}>\beta
\end{equation}
for a given $\alpha$. Nonetheless, there may be stronger restrictions on the sewn-in spacetime that save the area bound (\ref{eq:areab}) for $\beta$ outside the range of Eq. (\ref{eq:badbeta}).

\bibliographystyle{apsrev4-1long}
\bibliography{GeneralBibliography.bib}
\end{document}